\documentclass[a4paper,12pt]{article}
\usepackage{amsmath}
\usepackage{amsfonts}
\usepackage{amssymb}
\usepackage{latexsym}
\usepackage{epsfig}
\usepackage{graphicx}
\usepackage{oldgerm}
\usepackage{theorem}
\usepackage{bm}

\def\N{\mathbb{N}}
\def\C{\mathbb{C}}
\def\R{\mathbb{R}}
\def\H{\mathbb{H}}
\def\W{\mathbb{W}}
\def\P{\mathbb{P}}

\def\supp{{\rm supp}\,}
\def\rO{{\rm O}}
\def\rC{{\rm C}}

\def\sfV{{\sf V}}

\def\x{\bm{x}}
\def\X{\bm{X}}
\def\V{\bm{V}}
\def\B{\bm{B}}

\def\cK{{\cal K}}

\def\mM{\mathfrak{M}}

\theorembodyfont{\itshape}

\newtheorem{thm}{Theorem}[section]
\newtheorem{lem}[thm]{Lemma}

\newtheorem{prop}[thm]{Proposition}

\newcommand{\SSC}[1]{\section{#1}\setcounter{equation}{0}}
\newcommand{\qed}{\hbox{\rule[-2pt]{3pt}{6pt}}}

\begin{document}

\title{\bf 
Hydrodynamic Limit of Multiple SLE 
}
\author{
Ikkei Hotta
\footnote{
Department of Applied Science,
Yamaguchi University, 2-16-1 Tokiwadai,
Ube 755-8611, Japan;
e-mail: ihotta@yamaguchi-u.ac.jp
}, \, 
Makoto Katori
\footnote{
Department of Physics,
Faculty of Science and Engineering,
Chuo University, 
Kasuga, Bunkyo-ku, Tokyo 112-8551, Japan;
e-mail: katori@phys.chuo-u.ac.jp
}}
\date{8 March 2018}
\pagestyle{plain}
\maketitle

\begin{abstract}
Recently del Monaco and Schlei{\ss}inger addressed 
an interesting problem whether one can take the limit of
multiple Schramm--Loewner evolution (SLE) 
as the number of slits $N$ goes to infinity.
When the $N$ slits grow from points on the real line $\R$
in a simultaneous way and go to infinity
within the upper half plane $\H$,
an ordinary differential equation 
describing time evolution of the conformal map $g_t(z)$
was derived in the $N \to \infty$ limit, 
which is coupled with a complex Burgers equation
in the inviscid limit. 
It is well known that the complex Burgers equation
governs the hydrodynamic limit of the Dyson model
defined on $\R$ studied in random matrix theory, 
and when all particles start from
the origin, the solution of this Burgers equation is given
by the Stieltjes transformation of the measure which
follows a time-dependent version of Wigner's semicircle law.
In the present paper, first we study the hydrodynamic limit
of the multiple SLE in the case that all 
slits start from the origin. 
We show that the time-dependent version of Wigner's 
semicircle law determines the time evolution
of the SLE hull, $K_t \subset \H\cup \R$ ,
in this hydrodynamic limit. 
Next we consider the situation such that 
a half number of the slits start from $a>0$ and
another half of slits start from $-a < 0$,
and determine the multiple SLE in the 
hydrodynamic limit. 
After reporting these exact solutions,
we will discuss the universal long-term behavior of 
the multiple SLE and its hull $K_t$ in the hydrodynamic limit.

\vskip 0.2cm

\noindent{\bf Key words:} 
hydrodynamic limit; 
multiple Schramm--Loewner evolution (SLE); 
complex Burgers equation; 
Dyson model; 
Wigner's semicircle law

\end{abstract}
\vspace{3mm}

\SSC
{Introduction} \label{sec:introduction}

Construction and description of
stochastic interacting systems consisting of an infinite number of particles
have been important topics in probability theory
and nonequilibrium statistical mechanics
\cite{Lig85,Spo91,Lig99,For10,Kat15_Springer}.
In the present paper we report a trial to characterize
the infinite limit of stochastic interacting curves
in a plane.

\begin{figure}
\begin{center}
\includegraphics[width=0.6\textwidth]{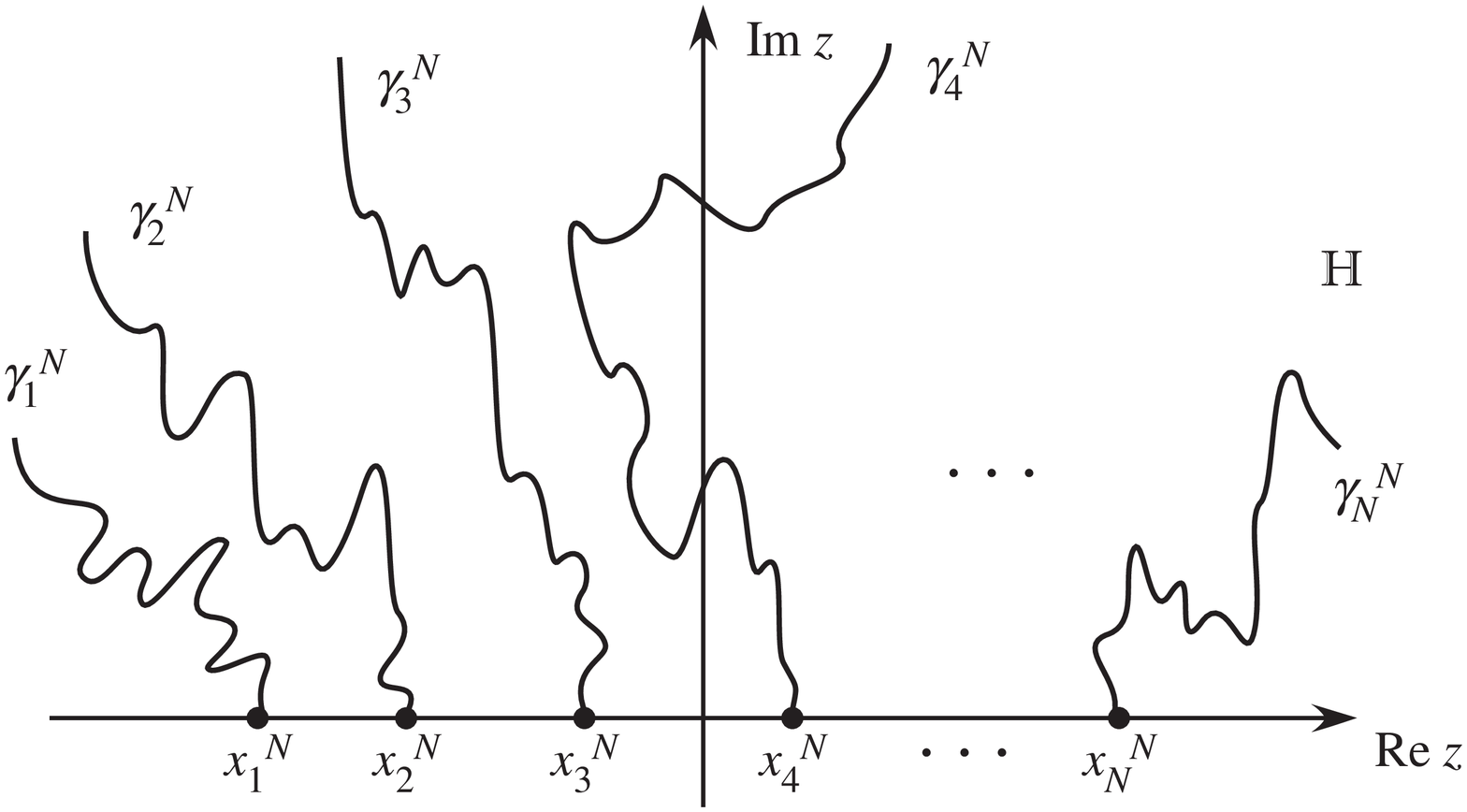}
\end{center}
\caption{
Schematic picture of $N$-tuples of non-intersecting slits
in $\H$,
$(\gamma^N_1, \gamma^N_2, \dots, \gamma^N_N)$, 
starting from $\x^N=(x^N_1, x^N_2, \dots, x^N_N) \in \W_N$.
}
\label{fig:mSLE1}
\end{figure}

Let $i=\sqrt{-1}$ and denote the upper half 
of the complex plane $\C$ as
\[
\H = \{z : z=x+i y, x \in \R, y > 0\}.
\]
For $N \in \N \equiv \{1,2, \dots\}$, consider the {\it Weyl chamber}
\[
\W_N =\{\x=(x_1, x_2, \dots, x_N) \in \R^N : x_1 < x_2 < \cdots < x_N \}.
\]
Given $\x^N=(x^N_1, \dots, x^N_N) \in \W_N$, we consider
$N$-tuples of slits in 
$\overline{\H} = \H \cup \R \cup \{\infty\}$ denoted as
$(\gamma^N_1, \gamma^N_2, \dots, \gamma^N_N)$, such that
$\gamma^N_j \in \overline{\H}$, $j=1,2, \dots, N$, are simple curves 
connecting $x_j$ and $\infty$, 
and they are non-intersecting, {\it i.e.},
\[
\gamma^N_j \cap \gamma^N_k = \emptyset,
\quad 1 \leq j < k \leq N.
\]
See Fig.\ref{fig:mSLE1}. 
Based on the theory of
{\it multiple Schramm--Loewner evolution (SLE)}
\cite{Car03,Car03c,Car04,BBK05,KL07,GMN16},
del Monaco and Schlei{\ss}inger \cite{dMS16}
have considered a one-parameter $(0 < \kappa \leq 4)$ 
family of probability laws $\P_{\kappa}^{\x^N, \infty}$
constructed from independent $N$ copies of
the one-dimensional Brownian motion as follows.

Let $B_j(t), t \geq 0$, $j=1,2, \dots, N$, be independent
one-dimensional standard Brownian motions.
The {\it Dyson model} with parameter $\beta >0$
studied in random matrix theory 
\cite{Dys62,Meh04,For10,AGZ10,Kat15_Springer}
is a system of stochastic differential equations (SDEs)
for the interacting particle system on $\R$, 
$\X^N(t)=(X^N_1(t), X^N_2(t), \dots, X^N_N(t))$,
\[
dX^N_j(t)=dB_j(t)+ \frac{\beta}{2} 
\sum_{\substack{ 1 \leq k \leq N, \cr k \not=j}}
\frac{dt}{X^N_j(t)-X^N_k(t)},
\quad t \in [0, T^{\x^N}],
\quad j=1,2, \dots, N,
\]
with an initial configuration $\X^N(0)=\x^N \in \W_N$,
and $T^{\x^N}=\inf \{t > 0 : \X^N(t) \notin \W_N\}$.
Here we set $\beta=8/\kappa \geq 2$ \cite{Car03,Car03c,Car04} and
perform a time change,
\[
V^N_j(t)=X^N_j(\kappa t/N), \quad
j=1,2, \dots, N.
\]
Then we have a system of SDEs for
$\V^N(t)=(V^N_1(t), \dots, V^N_N(t))$,
\begin{equation}
dV^N_j(t)= \sqrt{\frac{\kappa}{N}} dB_j(t)
+ \frac{1}{N} \sum_{\substack{1 \leq k \leq N, \cr k \not=j}}
\frac{4}{V^N_j(t)-V^N_k(t)} dt,
\quad t \in [0, \infty),
\quad j=1,2, \dots, N,
\label{eqn:V1}
\end{equation}
with the initial configuration
\[
\V^N(0)=\x^N \in \W_N.
\]
Here we have used the fact that $T^{\x^N}=\infty$,
$\forall \x^N \in \W_N$, with probability one
for the Dyson model with $\beta \geq 1$
\cite{RS93,GM13}.
With the solution $\V^N(t), t \in [0, \infty)$ of (\ref{eqn:V1}), 
the multiple SLE is introduced as
\begin{equation}
\frac{\partial g_t^N(z)}{\partial t}
= \frac{1}{N} \sum_{j=1}^N
\frac{2}{g_t^N(z)-V^N_j(t)}, \quad
t \geq 0,
\quad 
g_0(z)=z \in \H.
\label{eqn:mSLE1}
\end{equation}
Then, for $\kappa \in (0, 4]$, 
$\P_{\kappa}^{\x^N, \infty}$ is defined as 
the probability law of the $N$-tuples of slits
such that they are parameterized by 
$t \in [0, \infty]$ as $(\gamma^N_1(t), \dots, \gamma^N_N(t))$
with 
\[
\gamma^N_j(0)=x_j, \quad
\gamma^N_j(\infty)=\infty,
\quad j=1,2, \dots, N,
\]
and each realization of solution $g_t^N, t \in [0, \infty]$
for (\ref{eqn:mSLE1}) determines a time evolution
of the slits, 
$(\gamma^N_1(t), \dots, \gamma^N_N(t)), t \in [0, \infty]$,
in which $g_t^N$ is regarded as a time-dependent conformal
map (a {\it Loewner chain}) onto $\H$, and 
the domain of definition of $g_t^N$ is identified with 
$\H \setminus \bigcup_{j=1}^{N} \gamma^N_j(0, t]$
for each $t \in [0, \infty]$; 
\[
g_t^N : \mbox{conformal map} 
\quad 
\H \setminus \bigcup_{j=1}^{N} \gamma^N_j(0, t]
\to \H, \quad t \in [0, \infty], 
\]
where $\gamma^N_j(0, t] \equiv \bigcup_{0 < s \leq t} \gamma^N_j(s)$,
$j=1,2, \dots, N$.

Let $\mM$ be the space of probability measures on $\R$
equipped with its weak topology.
For $T > 0$, $\rC([0, T] \to \mM)$ denotes the
space of continuous processes defined in the time period $[0, T]$ 
realized in $\mM$.
Let $\delta_x(\cdot)$ be the Dirac measure centered at $x$;
$\delta_x(\{y\})=1$ if $y=x$ and $\delta_x(\{y\})=0$ otherwise.
We consider the empirical measure
of the solution $\V^N(t)$ of (\ref{eqn:V1}), 
\[
\sfV^N_t(\cdot)=\frac{1}{N} \sum_{j=1}^N \delta_{V^N_j(t)}(\cdot), \quad t \in [0, T], 
\]
as an element of $\rC([0, T] \to \mM)$,
whose initial value is given by
$\sfV_0^N(\cdot)=\frac{1}{N} \sum_{j=1}^N \delta_{x^N_j}(\cdot)$.

The following can be proved 
(see Proposition 4.3.10 in \cite{AGZ10}). 
\begin{prop}
\label{thm:AGZ1}
Assume that $(\x^N)_{N \in \N}$ is a sequence of initial configurations such that
$\x^N \in \overline{\W}_N$, 
\[
\sup_{N \geq 0} \frac{1}{N} \sum_{j=1}^N \log \{(x^N_j)^2+1\} < \infty,
\]
and $\sfV_0^N(\cdot)$ converges weakly to a measure $\mu_0(\cdot) \in \mM$
as $N \to \infty$. 
Then for any fixed $T < \infty$,
\[
(\sfV^N_t(\cdot))_{t \in [0, T]}
\to ^{\exists !} (\mu_t(\cdot))_{t \in [0, T]}
\quad \mbox{a.s. in $\rC([0, T] \to \mM)$},
\]
and the function defined by
\begin{equation}
M_t(z)=\int_{\R} \frac{2 \mu_t(du)}{z-u}
\label{eqn:Mt1}
\end{equation}
solves the equation
\begin{equation}
\frac{\partial M_t(z)}{\partial t} 
= -2 M_t(z) \frac{\partial M_t(z)}{\partial z},
\quad t \in [0, T], 
\quad z \in \C \setminus \R,
\label{eqn:Mt2}
\end{equation}
under the initial condition
\begin{equation}
M_0(z)= \int_{\R} \frac{2 \mu_0(du)}{z-u}.
\label{eqn:Mt0}
\end{equation}
\end{prop}
\vskip 0.3cm

The function $(M_t(z))_{t \in [0, T]}$, $z \in \C \setminus \R$ defined by 
the Stieltjes transformation (\ref{eqn:Mt1}) of $\mu_t(\cdot)$ is called
the {\it Green's function} (or the {\it resolvent}) 
for the measure-valued process $(\mu_t(\cdot))_{t \in [0, T]}$.
Equation (\ref{eqn:Mt2}) can be regarded as the
{\it complex Burgers equation in the inviscid limit} 
({\it i.e.}, the (complex) one-dimensional Euler equation).
Thus the $N \to \infty$ limit given by this theorem
is called the {\it hydrodynamic limit of the Dyson model} 
\cite{Cha92,RS93,BN08,AGZ10,BN10,FG15,AK16}. 
Note that the dependence on the parameter $\kappa \in (0, 4]$
disappears in the hydrodynamic limit.

Associated with this hydrodynamic limit of the Dyson model,
the following limit theorem was proved by
del Monaco and Schlei{\ss}inger.

\begin{thm}[del Monaco, Schlei{\ss}inger {\cite[Theorem 1.1]{dMS16}}]
\label{thm:dMS1}
Under the same assumption as given in Proposition \ref{thm:AGZ1},
in $N \to \infty$, 
$(g_t^N)_{N \in \N}$ converges locally uniformly in distribution  
to the solution $g_t$ of the deterministic Loewner equation
\begin{equation}
\frac{\partial g_t(z)}{\partial t}
= M_t(g_t(z)),
\quad t \geq 0, \quad
g_0(z)=z \in \H.
\label{eqn:dg}
\end{equation}
\end{thm}

Let $D_t, t \geq 0$ be the domain of definition of $g_t, t \geq 0$.
By the {\it Carath\'eodory kernel theorem} 
(see, for instance, Theorem 1.8 on page 29 in \cite{Pom75}),
the locally uniform convergence of $(g_t^N)_{N \in \N}$ to $g_t$ in $N \to \infty$
means that
\[
\H \setminus \bigcup_{j=1}^{N} \gamma_j^N(0, t] \to D_t, \quad t \geq 0, \quad
\mbox{in the sense of kernel convergence}.
\]
Moreover, del Monaco and Schlei{\ss}inger proved the following.
(See \cite{dMHS17} for the tightness results of the limit.) 

\begin{thm}[del Monaco, Schlei{\ss}inger {\cite[Theorem 1.2]{dMS16}}]
\label{thm:dMS2}
The set
$K_t \equiv \overline{\H \setminus D_t}$ is bounded for every $t \geq 0$ and 
there exists $T > 0$ such that for every $t > T$,
the boundary $\partial K_t \cap \H$ is an analytic curve in $\H$.
\end{thm}

In an earlier paper \cite{AK16}, the hydrodynamic limit of
the Dyson model was studied by solving the complex Burgers equation (\ref{eqn:Mt2})
explicitly for some special initial configurations $\mu_0$.
Here we regard the deterministic Loewner equation (\ref{eqn:dg}) 
for $g_t, t \geq 0$ as the {\it hydrodynamic limit of the multiple SLE},
and $K_t$ as the {\it SLE hull in the hydrodynamic limit}.
In the present paper first we characterize the
hydrodynamic limit of the multiple SLE by solving
(\ref{eqn:dg}) explicitly for two special initial configurations.
After reporting these exact solutions, a universal shape of
$K_t$ in long-term limit is discussed, which can be regarded as the
counterpart of the celebrated {\it Wigner's semicircle law}
realized in the hydrodynamic limit of the Dyson model 
\cite{Cha92,RS93,Meh04,For10,AGZ10,Kat15_Springer,AK16}.
The present study is an extension of the results
reported as examples and remarks in Sections 3.4 and 4 in \cite{dMS16}
and remarks in Section 2.5 in \cite{dMHS17}. 

The paper is organized as follows.
In Section \ref{sec:preliminaries} we explain a method to solve
the coupled system of the complex Burgers equation 
(\ref{eqn:Mt2}) and the multiple SLE in the hydrodynamic limit (\ref{eqn:dg}).
This method was briefly mentioned in Remark 3.11 in \cite{dMS16}.
Section \ref{sec:exact1} is devoted to reporting the exact results
for the system starting from a single source at the origin,
$\mu_0=\delta_0$.
We can see the same statement as Proposition \ref{thm:SLE_gt_single}
and the same figure as Fig.\ref{fig:hydro_SLE_single1} 
in Section 4 of \cite{dMS16},
but the explicit expression for the SLE hull 
(\ref{eqn:SLEhull_single1}) with (\ref{eqn:SLEhull_universal}) is
first given in Proposition \ref{thm:SLEhull_single} in the present paper.
New exact results are reported in Section \ref{sec:exact2}
for the system starting from the two sources.
The long-term asymptotics of the SLE and its hull
$K_t$ are generally discussed in Section \ref{sec:asym}.
Concluding remarks are given in Section \ref{sec:remarks}.

\SSC
{Preliminaries} \label{sec:preliminaries}

\subsection{Transformation from $g_t$ to $h_t$} \label{sec:ht}
In Remark 3.11 in \cite{dMS16}, the following 
transformation $g_t \to h_t$ was introduced, 
\begin{equation}
g_t(z)=h_t(z)+ 2 t M_t(g_t(z)), 
\label{eqn:ht1}
\end{equation}
where $h_t$ is chosen as the following equalities hold,
\begin{align}
M_t(g_t(z))=& M_0(h_t(z)), \quad t \geq 0, 
\nonumber\\
g_0(z) =& h_0(z)= z \in \H.
\label{eqn:ht2}
\end{align}
The compatibility of (\ref{eqn:ht1}) and (\ref{eqn:ht2}) is
guaranteed by the fact that the solution $M_t(z)$ of the
complex Burgers equation (\ref{eqn:Mt2}) satisfies the 
functional equation
\[
M_t(z)=M_0(z-2t M_t(z)).
\]
See, for instance, Theorem 1.2 (ii) in \cite{AK16}.

Then the following is verified.
\begin{lem}[del Monaco, Schlei{\ss}inger {\cite[Remark 3.11]{dMS16}}]
\label{thm:ht_eq}
Assume that 
the function $h_t(z)$ solves
the following partial differential equation,
\begin{equation}
\frac{\partial h_t(z)}{\partial t}
= - \frac{M_0(h_t(z))}{\displaystyle{1+2t \frac{\partial M_0}{\partial z}(h_t(z))}},
\quad h_0(z)=z.
\label{eqn:ht_eq1}
\end{equation}
Then $g_t(z)$ is obtained from $h_t(z)$ by the relation
\begin{equation}
g_t(z)=h_t(z)+ 2 t M_0(h_t(z)). 
\label{eqn:ht4}
\end{equation}
\end{lem}
\vskip 0.3cm
\noindent{\it Proof.} \,
By the definition (\ref{eqn:ht1}) of $h_t$
and the condition (\ref{eqn:ht2}) for $h_t$,
(\ref{eqn:ht4}) is concluded. 
By differentiating (\ref{eqn:ht4}) by $t$, we obtain
\[
\frac{\partial g_t(z)}{\partial t}
= \frac{\partial h_t(z)}{\partial t}
+ 2 M_0(h_t(z)) + 2 t \frac{\partial M_0}{\partial z}(h_t(z))
\frac{\partial h_t(z)}{\partial t}.
\]
On the other hand, by the deterministic Loewner equation (\ref{eqn:dg})
and the equality (\ref{eqn:ht2}), 
\[
\frac{\partial g_t(z)}{\partial t}
= M_t(g_t(z))=M_0(h_t(z)).
\]
Then we have the equation
\[
\frac{\partial h_t(z)}{\partial t}
+ 2 M_0(h_t(z)) + 2 t \frac{\partial M_0}{\partial z}(h_t(z))
\frac{\partial h_t(z)}{\partial t}
=M_0(h_t(z)),
\]
which is equivalent with (\ref{eqn:ht_eq1}).
Then the proof is completed. \qed
\vskip 0.3cm

\subsection{SLE hull $K_t$ in the hydrodynamic limit} \label{sec:Kt}

For each $t \in [0, \infty)$, the boundary of SLE hull
in the hydrodynamic limit, $\partial K_t \subset \H \cup \R$,
is given by the inverse image $g_t^{-1}$ 
of the closed support of $\mu_t$;
\begin{equation}
\partial K_t= g_t^{-1}(\supp \mu_t), \quad t \in [0, \infty),
\label{eqn:Kt1}
\end{equation}
and the SLE hull in the hydrodynamic limit is given by
\begin{equation}
K_t= \bigcup_{0 \leq s \leq t} \partial K_s,
\quad t \in [0, \infty). 
\label{eqn:Kt2}
\end{equation}

We assume that there is a set $I \subset \R$ and 
the elements of $\supp \mu_t$ are
parameterized by $\xi \in I$ as
\begin{equation}
\supp \mu_t = \{ \sigma_t(\xi) : \xi \in I \}, \quad
t \in [0, \infty).
\label{eqn:parameter1}
\end{equation}
Define
\[
\Gamma_t(\xi) = g_t^{-1}(\sigma_t(\xi)),
\quad \xi \in I, \quad t \in [0, \infty).
\]
Then (\ref{eqn:Kt1}) gives
\[
\partial K_t = \{ \Gamma_t(\xi) : \xi \in I \}, \quad
t \in [0, \infty).
\]
By (\ref{eqn:ht4}), we have the equality
\begin{equation}
\sigma_t(\xi)=h_t(\Gamma_t(\xi))+ 2 t M_0(h_t(\Gamma_t(\xi))),
\quad t \in [0, \infty).
\label{eqn:parameter3}
\end{equation}

\SSC
{Exact Results for the System Starting from Single Source} \label{sec:exact1}
\subsection{SLE $g_t$ in the hydrodynamic limit} \label{sec:single_source_SLE}
Consider the multiple SLE in the case that
all slits start from the origin, which is obtained by
taking the limit $x^N_j \to 0$, $1 \leq \forall j \leq N$
for $\x^N=(x^N_1, x^N_2, \dots, x^N_N) \in \W$. 
In the hydrodynamic limit $N \to \infty$, 
this situation is realized by setting
\begin{equation}
\mu_0=\delta_0, \quad
\mbox{that is,} \quad
\mu_0(du) = \delta(u) du,
\label{eqn:single_0}
\end{equation}
where $\delta(u)$ denotes Dirac's delta function.
We say in this situation that
the multiple SLE in the hydrodynamic limit 
starts from a single source located at the origin.

In this case, (\ref{eqn:Mt0}) is given by
\begin{equation}
M_0(z)=\int_{\R} \frac{2 \delta_0(du)}{z-u}=\frac{2}{z}.
\label{eqn:M0_single}
\end{equation}
Under this initial condition the complex Burgers equation (\ref{eqn:Mt2}) 
is solved as
\begin{align}
M_t(z) =& \frac{1}{4t}(z-\sqrt{z^2-16 t})
\nonumber\\
=& \frac{4}{z+\sqrt{z^2-16 t}}, \quad
t \in [0, \infty). 
\label{eqn:Mt_single}
\end{align}
Then the hydrodynamic limit of the multiple SLE (\ref{eqn:dg})
which we want to solve now is given by
\begin{equation}
\frac{\partial g_t(z)}{\partial t}
= \frac{4}{g_t(z)+\sqrt{g_t(z)^2-16t}}, \quad
t \geq 0, \quad g_0(z)=z \in \H.
\label{eqn:SLE_single}
\end{equation}
Thanks to Lemma \ref{thm:ht_eq}, 
however, we do not need to solve directly this equation
(\ref{eqn:SLE_single}) to determine $g_t$, $t \in [0, \infty)$. 
Only using the simple initial condition $M_0(z)$ given by (\ref{eqn:M0_single}),
we can obtain Eq.(\ref{eqn:ht_eq1}) for $h_t(z)$,
and through (\ref{eqn:ht4}) $g_t(z)$ is determined.
In the present case, Eq.(\ref{eqn:ht_eq1}) becomes
\begin{align}
\label{eqn:ht_eq_single1}
\frac{\partial h_t(z)}{\partial t}
=& - \frac{\displaystyle{ \frac{2}{h_t(z)} }}
{\displaystyle{1- \frac{4t}{h_t(z)^2}}},
\\
\label{eqn:ht_eq_single2}
h_0(z) =& z \in \H.
\end{align}
It is easy to verify that Eq.(\ref{eqn:ht_eq_single1}) is equivalent with
\[
\frac{\partial}{\partial t} \log h_t(z) 
=\frac{\partial}{\partial t} \left( - \frac{2t}{h_t(z)^2} \right).
\]
Then we have 
$h_t(z) = c_1 e^{-2t/h_t(z)^2}$,
where $c_1$ is an integral constant.
By the initial condition (\ref{eqn:ht_eq_single2}),
the constant is determined as $c_1=z$.
We rewrite the obtained equation as
\begin{equation}
-\frac{4t}{h_t(z)^2} e^{-4t/h_t(z)^2}
= - \frac{4t}{z^2}.
\label{eqn:eq_s_A1}
\end{equation}

Here we consider the {\it Lambert $W$ function} 
(see, for instance, \cite{CGHJK96,Veb12}).
This function is defined as the inverse function
of the mapping
\[
x \mapsto x e^{x}.
\]
This mapping is not injective, and the Lambert $W$
function has two real branches with 
a branching point at $(-e^{-1}, -1)$ 
in the real plane $(x, W) \in \R^2$.
We take the upper branch $W_0(x)$ defined
for $x \in [-e^{-1}, \infty)$.
By this definition, we can show that
\begin{align*}
&W_0(x) e^{W_0(x)} = x,
\nonumber\\
&W_0(0) = 0, \quad W_0(e)=1,
\end{align*}
and
\begin{equation}
W_0(x) \simeq x \quad
\mbox{in $x \to 0$}.
\label{eqn:Lambert1}
\end{equation}
Define the complex function $W_0(z), z \in \C$ as 
an analytic continuation of $W_0(x) \in [-e^{-1}, \infty) \subset \R$.
$W_0(z)$ is analytic at $z=0$ having the expansion
\[
W_0(z)=\sum_{n=1}^{\infty} \frac{(-n)^{n-1}}{n!} z^n,
\]
whose radius of convergence is $e^{-1}$.
The branch cut of $W_0(z)$ on the complex plane $\C$ is given by
$(-\infty, -e^{-1}] \subset \R$.
See \cite{CGHJK96} for more details.

Equation (\ref{eqn:eq_s_A1}) is now rewritten as
\[
W_0 \left( - \frac{4t}{z^2} \right)
=-\frac{4t}{h_t(z)^2},
\]
which gives
\begin{equation}
h_t(z)= \pm i \sqrt{\frac{4t}{W_0(-4t/z^2)}}
=\pm 2i \sqrt{t} \frac{1}{\sqrt{W_0(-4t/z^2)}}.
\label{eqn:eq_s_A3}
\end{equation}
If we take the limit $t \to 0$ in (\ref{eqn:eq_s_A3}),
we have
$h_0(z) =\pm i \sqrt{-z^2}$,
where (\ref{eqn:Lambert1}) was used.
We choose the square root branch as
$\sqrt{-z^2}=-i z, z \in \H$, and hence
$h_0(z)=\pm z$.
Due to the initial condition (\ref{eqn:ht_eq_single2}) of $h_t$,
we should choose the plus sign, and we obtain
\begin{equation}
h_t(z)= 2i \sqrt{t}  \frac{1}{\sqrt{W_0(-4t/z^2)}}.
\label{eqn:ht_single_sol1}
\end{equation}

From this solution $h_t(z)$, we obtain 
the solution $g_t$ of (\ref{eqn:SLE_single}) by (\ref{eqn:ht4})
following Lemma \ref{thm:ht_eq}.
By (\ref{eqn:M0_single}), Eq.(\ref{eqn:ht4}) becomes
\begin{equation}
g_t(z)=h_t(z)+\frac{4t}{h_t(z)}.
\label{eqn:gt_single1}
\end{equation}
Insert (\ref{eqn:ht_single_sol1}) into (\ref{eqn:gt_single1}), we obtain
the exact solution of Eq.(\ref{eqn:SLE_single}) 
as following.

\begin{prop}
\label{thm:SLE_gt_single}
The exact solution of the multiple SLE in the hydrodynamic limit
starting from the single source (\ref{eqn:single_0}) is given by
\[
g_t(z) = 2 i \sqrt{t} \left\{
\frac{1}{\sqrt{W_0(-4t/z^2)}}
- \sqrt{W_0(-4t/z^2)} \right\},
\]
where $W_0$ is the Lambert $W$ function 
satisfying (\ref{eqn:Lambert1}). 
\end{prop}

\subsection{SLE hull $K_t$ in the hydrodynamic limit} \label{sec:single_source_SLEhull}

Now we determine the SLE hull by (\ref{eqn:Kt1}) and (\ref{eqn:Kt2}). 
So far we did not use the explicit expression of the
solution (\ref{eqn:Mt_single}), but now we have to know
the support of $\mu_t$ in the expression (\ref{eqn:Mt1})
of the solution (\ref{eqn:Mt_single}). 
By the same argument as given in Section 2 of \cite{AK16},
the following equality can be derived,
\[
\mu_t(du)=- \Im
\left[ \lim_{\varepsilon \downarrow 0} 
\frac{1}{2 \pi} M_t(u+ i \varepsilon) \right] du,
\quad t \in [0,\infty), \quad u \in \R.
\]
From (\ref{eqn:Mt_single}), we have 
\begin{align}
\mu_t(du) 
&= \begin{cases}
\displaystyle{ 
\frac{1}{8 \pi t} \sqrt{16 t-u^2} du
},
& \quad
\mbox{if $|u| < 4 \sqrt{t}$},
\\
0,
& \quad
\mbox{if $|u| \geq 4 \sqrt{t}$},
\end{cases}
\label{eqn:Wigner}
\end{align}
for $t \in [0, \infty)$.
This is a {\it time-dependent version of 
Wigner's semicircle law} 
in the present time change \cite{Cha92,RS93} 
(see also Section 3.9 of \cite{Kat15_Springer}).
Therefore, we can conclude that 
$\supp \mu_t = [-4 \sqrt{t}, 4 \sqrt{t} ]$.

The parameterization of $\supp \mu_t$ assumed as (\ref{eqn:parameter1})
is now realized as
\begin{equation}
\sigma_t(\xi) =4 \sqrt{t} \xi, \quad
\xi \in I \equiv [-1, 1].
\label{eqn:parameter_single1}
\end{equation}
By (\ref{eqn:M0_single}), Eq.(\ref{eqn:parameter3}) is given as
\begin{equation}
\sigma_t(\xi)=h_t(\Gamma_t(\xi))+ \frac{4t}{h_t(\Gamma_t(\xi))},
\quad t \in [0, \infty).
\label{eqn:eq_single1}
\end{equation}
Put 
\[ h_t(\Gamma_t(\xi))=v+iw, \quad
v=v(\sigma_t(\xi),t) \in \R, \quad w=w(\sigma_t(\xi), t) \in \R.
\]
Since $\sigma_t(\xi) \in \R$, Eq.(\ref{eqn:eq_single1}) gives
\[
v^2-w^2-\sigma_t(\xi) v + 4t =0,
\quad
2v = \sigma_t(\xi).
\]
They are solved as
\[
v=\frac{\sigma_t(\xi)}{2}, \quad
w=\frac{1}{2} \sqrt{16t -\sigma_t(\xi)^2}, 
\]
and we obtain 
\begin{equation}
h_t(\Gamma_t(\xi))=
\frac{1}{2} \left\{ \sigma_t(\xi) + i \sqrt{16 t-\sigma_t(\xi)^2} \right\}.
\label{eqn:ht_single_sol2}
\end{equation}

On the other hand, the solution (\ref{eqn:ht_single_sol1}) gives
the equality,
\[
W_0 \left( - \frac{4t}{\Gamma_t(\xi)^2} \right)
= - \frac{4t}{h_t(\Gamma_t(\xi))^2},
\quad t \in [0, \infty).
\]
By (\ref{eqn:ht_single_sol2}), it gives
\[
W_0 \left( - \frac{4t}{\Gamma_t(\xi)^2} \right)
=1-\frac{\sigma_t(\xi)^2}{8t}
+ i \frac{\sigma_t(\xi)}{8t} 
\sqrt{16t-\sigma_t(\xi)^2}.
\]
Since the Lambert function $W_0$ is defined as
the inverse function of the mapping 
$x \mapsto x e^x$, the above equation is equivalent with
\begin{align*}
-\frac{4t}{\Gamma_t(\xi)^2}
 =& \left( 1- \frac{\sigma_t(\xi)^2}{8t}
+ i \frac{\sigma_t(\xi)}{8t} \sqrt{16 t - \sigma_t(\xi)^2} \right)
\nonumber\\
& \times
\exp \left( 1- \frac{\sigma_t(\xi)^2}{8t}
+ i \frac{\sigma_t(\xi)}{8t} \sqrt{16 t - \sigma_t(\xi)^2} \right),
\end{align*}
which gives
\begin{align}
\Gamma_t(\xi)^2
=& -4t \left( 1- \frac{\sigma_t(\xi)^2}{8t}
- i \frac{\sigma_t(\xi)}{8t} \sqrt{16 t - \sigma_t(\xi)^2} \right)
\nonumber\\
& \times
\exp \left(-1+ \frac{\sigma_t(\xi)^2}{8t}
- i \frac{\sigma_t(\xi)}{8t} \sqrt{16 t - \sigma_t(\xi)^2} \right).
\label{eqn:Gamma_single_sol1}
\end{align}
Here we use the parameterization (\ref{eqn:parameter_single1}).
Moreover, we put
\begin{equation}
\xi=\sin \varphi, \quad \varphi \in [-\pi/2, \pi/2].
\label{eqn:parameter_varphi1}
\end{equation}
Then we see that (\ref{eqn:Gamma_single_sol1}) is
simplified and we obtain
\[
\widetilde{\Gamma}_t(\varphi)
\equiv \Gamma_t(\sin \varphi)
=2 i \sqrt{t} \exp \left( - i \varphi - \frac{e^{2 i \varphi}}{2} \right).
\]

The above results are summarized as follows.

\begin{prop}
\label{thm:SLEhull_single}
Assume that the multiple SLE in the hydrodynamic limit
starts from a single source at the origin, $\mu_0=\delta_0$.
Then the SLE hull is given by 
\begin{equation}
K_t= \sqrt{t} \cK, \quad t \in [0, \infty),
\label{eqn:SLEhull_single1}
\end{equation}
with
\begin{equation}
\cK = K_1= \left\{ \widetilde{\Gamma}_s(\varphi):
\widetilde{\Gamma}_s(\varphi)
= 2 i r \exp \left( - i \varphi - \frac{e^{2 i \varphi}}{2} \right), 
-\frac{\pi}{2} \leq \varphi \leq \frac{\pi}{2}, 
0 \leq r \leq 1 \right\}.
\label{eqn:SLEhull_universal}
\end{equation}
\end{prop}
\vskip 1cm

\begin{figure}
\begin{center}
\includegraphics[width=1.0\textwidth]{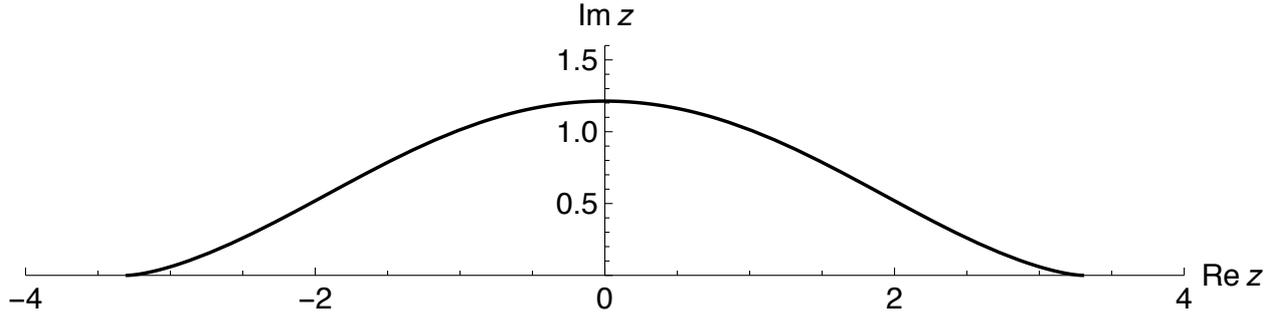}
\end{center}
\caption{The curve shows the boundary $\partial \cK$
of (\ref{eqn:SLEhull_universal}).
Here $\cK \cap \R =[-2 \sqrt{e}, 2 \sqrt{e}]=[-3.297 \cdots, 3.297 \cdots]$
and
$\max_{z \in \cK} \Im z =2/\sqrt{e}=1.213 \cdots$.
}
\label{fig:hydro_SLE_single1}
\end{figure}

It is easy to verify that
\begin{align}
\Re \widetilde{\Gamma}_t(-\varphi) =& - \Re \widetilde{\Gamma}_t(\varphi),
\quad
\Im \widetilde{\Gamma}_t(-\varphi)= \Im \widetilde{\Gamma}_t(\varphi),
\quad -\frac{\pi}{2} \leq \varphi \leq \frac{\pi}{2},
\nonumber\\
\max_{\varphi \in [-\pi/2, \pi/2]} \Im \widetilde{\Gamma}_t(\varphi) =&
\Im \widetilde{\Gamma}_t(0) = 2 \sqrt{\frac{t}{e}},
\nonumber\\
\widetilde{\Gamma}_t( \pm (\pi/2-\varepsilon))
=& \pm 2 \sqrt{t e} (1-\varepsilon^2)+ i \frac{4 \sqrt{te}}{3} \varepsilon^3
+\rO(\varepsilon^4), \quad \varepsilon >0.
\label{eqn:edges1a}
\end{align}
The boundary $\partial \cK = \partial K_1$  
is shown by a curve in Fig.\ref{fig:hydro_SLE_single1}.
Remark that the interval
$K_t \cap \R = \sqrt{t} \cK \cap \R
=[-2 \sqrt{t e}, 2 \sqrt{t e}]$
corresponds to the branch cut of $W_0(-4t/z^2)$,
which is considered as a complex function of $z$
for each time $t >0$ in (\ref{eqn:ht_single_sol1});
$-4t/z^2 \in (-\infty, -e^{-1}] 
\Longleftrightarrow |z| \leq 2 \sqrt{t e}$, $z \in \R$.

The expansion (\ref{eqn:edges1a})
implies that in the vicinity of the edges $\pm 2 \sqrt{te} \in \R$,
the boundary
$\{ z = x+i y : z \in \partial K_t =\sqrt{t} \partial \cK \}$ behaves as
\begin{equation}
y \simeq \begin{cases}
\displaystyle{ \frac{\sqrt{2}}{3} (t e)^{-1/4} 
(2 \sqrt{t e} - x)^{3/2} },
\quad & \mbox{if $x \leq 2 \sqrt{t e}$},
\\
\displaystyle{ \frac{\sqrt{2}}{3} (t e)^{-1/4} 
(x+2 \sqrt{t e})^{3/2} },
\quad & \mbox{if $x \geq - 2 \sqrt{t e}$}.
\end{cases}
\label{eqn:edges_single}
\end{equation}

\SSC
{Exact Results for the System Starting from Two Sources} \label{sec:exact2}
\subsection{SLE $g_t$ in the hydrodynamic limit} \label{sec:two_source_SLE}
Next we consider the multiple SLE in the case that
a half number of slits start from $x=a >0$ and 
another half of slits start from $x=-a < 0$, 
which is obtained for an even $N$ by
taking the limit $x^N_j \to -a$, $1 \leq j \leq N/2$,
and $x^N_j \to a$, $N/2+1 \leq j \leq N$
for $\x^N=(x^N_1, x^N_2, \dots, x^N_N) \in \W$. 
In the hydrodynamic limit $N \to \infty$, 
this situation is realized by setting 
\begin{equation}
\mu_0= \frac{1}{2} (\delta_a + \delta_{-a} ), \quad
\mbox{that is}, \quad
\mu_0(du) = \frac{1}{2} \{ \delta(u-a)+\delta(u+a) \} du.
\label{eqn:double_0}
\end{equation}
We say in this situation that
the multiple SLE in the hydrodynamic limit
starts from the two sources. 

In this case, the initial condition of
the complex Burgers equation is given by
\begin{align}
M_0(z) &=\int_{\R} \frac{2 \delta_0(du)}{z-u}
\nonumber\\
&=\frac{1}{z-a} + \frac{1}{z+a}
= \frac{2z}{z^2-a^2}.
\label{eqn:M0_double}
\end{align}
Equation (\ref{eqn:ht_eq1}) for $h_t(z)$ 
given in Lemma \ref{thm:ht_eq} is then
\begin{align}
\frac{\partial h_t(z)}{\partial t}
&= - \frac{\displaystyle{ \frac{1}{h_t(z)-a} + \frac{1}{h_t(z)+a} }}
{\displaystyle{1- 2t \left\{ \frac{1}{(h_t(z)-a)^2} + \frac{1}{(h_t(z)+a)^2} \right\} }}
\nonumber\\
\label{eqn:ht_eq_double1}
&= - \frac{ \displaystyle{
\frac{2 h_t(z)}{h_t(z)^2-a^2} }}
{\displaystyle{
1- 4t \frac{h_t(z)^2+a^2}{(h_t(z)^2-a^2)^2} }},
\\
\label{eqn:ht_eq_double2}
h_0(z) &= z \in \H.
\end{align}
Equation (\ref{eqn:ht4}) in Lemma \ref{thm:ht_eq}
becomes 
\begin{align}
g_t(z) &= h_t(z)+2t \left\{ \frac{1}{h_t(z)-a} + \frac{1}{h_t(z)+a} \right\}
\nonumber\\
&= h_t(z) + \frac{ 4t h_t(z)}{h_t(z)^2-a^2}.
\label{eqn:gt_double1}
\end{align}
We have found that (\ref{eqn:ht_eq_double1}) is equivalent with
\[
\frac{\partial}{\partial t}
\log \sqrt{h_t(z)^2-a^2}
= \frac{\partial}{\partial t}
\left( - \frac{2t h_t(z)^2}{(h_t(z)^2-a^2)^2} \right).
\]
This is solved as
$\sqrt{h_t(z)^2-a^2}
= c_2 e^{-2t h_t(z)^2/(h_t(z)^2-a^2)^2}$
with an integral constant $c_2$.
By the initial condition (\ref{eqn:ht_eq_double2}), 
it is determined as $c_2=\sqrt{z^2-a^2}$, and 
we obtain the equation
\[
h_t(z)^2-a^2
=(z^2-a^2)
e^{- 4 t h_t(z)^2/(h_t(z)^2-a^2)^2}.
\]
We solve this equation for $z$ as
\[
z=\sqrt{a^2+(h_t(z)^2-a^2) e^{4t h_t(z)^2/(h_t(z)^2-a^2)^2} },
\]
which satisfies the initial condition (\ref{eqn:ht_eq_double2});
$h_0(z)=z$.
Define a function
\begin{equation}
V_t(z)=\sqrt{1+(z^2-1) e^{4 tz^2/(z^2-1)^2} },
\quad t \in [0, \infty), \quad z \in \H.
\label{eqn:Vt1}
\end{equation}
Then the solution $h_t(z)$ of (\ref{eqn:ht_eq_double1}) under
(\ref{eqn:ht_eq_double2}) is expressed as
\begin{equation}
h_t(z)=a V_{t/a^2}^{-1} \left( \frac{z}{a} \right),
\label{eqn:ht_double_sol1}
\end{equation}
where $V_t^{-1}$ denotes the inverse function of $V_t$
for each $t \in [0, \infty)$.

Insert (\ref{eqn:ht_double_sol1}) into (\ref{eqn:gt_double1}), we obtain
the exact solution $g_t$
as following.

\begin{prop}
\label{thm:SLE_gt_double}
The exact solution of the multiple SLE in the hydrodynamic limit
starting from the two sources (\ref{eqn:double_0}) is given by
\begin{equation}
g_t(z)=a \left\{ V_{t/a^2}^{-1} \left( \frac{z}{a} \right)
+ \frac{4 (t/a^2) V_{t/a^2}^{-1}(z/a)}{(V_{t/a^2}^{-1}(z/a))^2-1} \right\},
\quad t \in [0, \infty),
\label{eqn:gt_double_sol1}
\end{equation}
where $V_t^{-1}$ is the inverse function of $V_t$
defined by (\ref{eqn:Vt1}).
\end{prop}

\subsection{SLE hull $K_t$ in the hydrodynamic limit} \label{sec:two_source_SLEhull}
The complex Burgers equation (\ref{eqn:Mt2}) was solved
under the two-source initial condition (\ref{eqn:double_0}) 
in Section 6.5 of \cite{Nadal11}, in Section 2.3 of \cite{Warchol14},
and in Section 4.2 of \cite{AK16}.
See also \cite{BGNW15}.
The critical time is given by
\[
t_{\rm c}=t_{\rm c}(a)= \frac{a^2}{4},
\]
and the support of $\mu_t, t \geq 0$ is determined as
\begin{equation}
\supp \mu_t = 
\begin{cases}
\{x \in \R : a b_{-}(t/a^2) \leq |x| \leq a b_{+}(t/a^2) \},
\quad & \mbox{if $0 \leq t < t_{\rm c}$},
\\
\{x \in \R : 0 \leq |x| \leq a b_{+}(t/a^2) \},
\quad & \mbox{if $t \geq t_{\rm c}$}, 
\end{cases}
\label{eqn:supp_d1}
\end{equation}
where
\begin{equation}
b_{\pm}(t) = \sqrt{
(1+2t) \pm 2 \sqrt{t(t+2)} }
\left\{ 
(1-t) \pm \sqrt{t(t+2)} \right\}.
\label{eqn:bpm1}
\end{equation}

It is easy to verify that
\[
b_{\pm}(t)^2 
=1+10t - 2 t^2 \pm 2 (2+t) \sqrt{t(2+t)}
\]
and
\[
b_+(t) b_-(t)=(1-4t)^{3/2}.
\]
Then (\ref{eqn:bpm1}) is written as
\begin{align}
b_-(t) =& \sqrt{1+10t-2t^2-2(2+t)\sqrt{t(2+t)} }
\nonumber\\
=& \frac{(1-4t)^{3/2}}{\sqrt{1+10t-2t^2+2(2+t)\sqrt{t(2+t)}}},
\nonumber\\
b_+(t) =& \sqrt{1+10t-2t^2+2(2+t) \sqrt{t(2+t)}}.
\label{eqn:bpm1b}
\end{align}

Since the system is symmetric with respect to
the imaginary axis in $\C$, we need to consider only
the non-negative part of the support
\[
\supp \mu_t^+ = \{u \in \supp \mu_t : u \geq 0\},
\quad t \geq 0.
\]
Moreover, the above formulas (\ref{eqn:gt_double_sol1})
and (\ref{eqn:supp_d1}) implies that
we can set $a=1$ in calculation, since
the system obeys the diffusion scaling and hence 
the general solution will be obtained
by just setting $t \to t/a^2, z \to z/a$
for $a > 0$.
The parameterization for $\supp \mu_t^+$ is given by
\begin{equation}
\sigma_t^+(\xi)=
\begin{cases}
(1-\xi) b_{-}(t)+ \xi b_{+}(t),
\quad & \mbox{if $0 \leq t <1/4$},
\cr
\xi b_{+}(t),
\quad & \mbox{if $t \geq 1/4$},
\end{cases}
\quad \xi \in I^+ \equiv [0, 1].
\label{eqn:parameter_double_1}
\end{equation}

By (\ref{eqn:M0_double}), Eq.(\ref{eqn:parameter3}) is given as
\begin{equation}
\sigma_t^+(\xi) 
= h_t(\Gamma_t^+(\xi))
+ \frac{4 t h_t(\Gamma_t^+(\xi))}{h_t(\Gamma_t^{+}(\xi))^2-1}, \quad t \in [0, \infty).
\label{eqn:ht_eq_double3}
\end{equation}
Put
\[ h_t(\Gamma_t^+(\xi))=v+iw, \quad
v=v(\sigma_t^+(\xi),t) \in \R, \quad w=w(\sigma_t^+(\xi), t) \in \R.
\]
Since $\sigma_t^+(\xi) \geq 0$, Eq.(\ref{eqn:ht_eq_double3}) gives
\begin{align}
& v^3-\sigma_t^+(\xi) v^2 - \{ 3 w^2 -(4t-1) \} v+ \sigma_t^+(\xi) (w^2+1)=0,
\nonumber\\
& w \{ w^2+2 \sigma_t^+(\xi) v -3v^2-(4t-1) \} =0.
\label{eqn:xy_double1}
\end{align}
We need the solution such that $w$ is not identically zero.
Then we have chosen the solution of (\ref{eqn:xy_double1}) as
\begin{align}
& v=v(\sigma_t^+(\xi), t)
\equiv -\frac{(S(\sigma_t^+(\xi), t) - \sigma_t^+(\xi))^2-3(4t-1)}{6 S(\sigma_t^+(\xi), t)},
\nonumber\\
& w=w(\sigma_t^+(\xi), t)
\equiv \sqrt{3 v(\sigma_t^+(\xi), t)^2
- 2 \sigma_t^+(\xi) v(\sigma_t^+(\xi), t)+4t-1},
\quad t \in [0, \infty), 
\label{eqn:xy_double_sol1}
\end{align}
where
\begin{equation}
S(\sigma, t)= 
\sqrt[3]{\sigma^{3}+9 \sigma^{2} 
\sqrt{- \sigma^{2}-2(2t^2-10 t -1) + 3 (4t-1)^3}-9 \sigma(2 t+1) }.
\label{eqn:xy_double_sol2}
\end{equation}
On the other hand, the solution (\ref{eqn:ht_double_sol1})
gives
\[
\Gamma_t^+(\xi) =V_t(h_t(\Gamma_t^+(\xi))), 
\quad t \in [0, \infty).
\]

The results are summarized as follows.

\begin{prop}
\label{thm:SLEhull_double}
Assume that the multiple SLE in the hydrodynamic limit
starts from the two sources 
$\mu_0=(\delta_a+\delta_{-a})/2$, $a>0$.
Then the SLE hull is given by 
\[
K_t= K_t^+ \cup K_t^-, \quad t \in [0, \infty),
\]
with
\begin{align*}
K_t^+ =& \Bigg\{ \Gamma_s^+(\xi):
\Gamma_s^+(\xi)
=a V_{s/a^2} \Bigg(
a \Big\{ v(\sigma_{s/a^2}^+(\xi), s/a^2) + i w(\sigma_{s/a^2}^+(\xi), s/a^2) \Big\} \Bigg), 
\nonumber\\
& \hskip 9cm 
\quad
\xi \in [0, 1], \,
0 \leq s \leq t \Bigg\},
\nonumber\\
K_t^- =& - \overline{K_t^+}
\equiv \{ - \overline{z} =-x+i y : z=x+ i y \in  K_t^+ \}, 
\end{align*}
where $v(\sigma_t^+(\xi), t)$ and
$w(\sigma_t^+(\xi), t)$ are given by
(\ref{eqn:xy_double_sol1}) 
with (\ref{eqn:bpm1}), (\ref{eqn:parameter_double_1})
and (\ref{eqn:xy_double_sol2}). 
\end{prop}

\begin{figure}
\begin{center}
\includegraphics[width=1.0\textwidth]{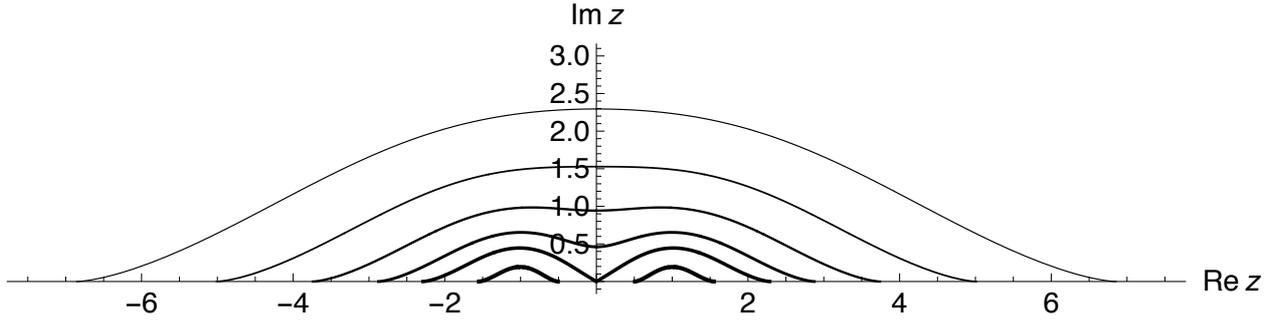}
\end{center}
\caption{The boundaries of the SLE hull
in the hydrodynamic limit, 
$\partial K_t$, are shown 
for $t=0.05$ (the thickest curve), $0.25, 0.5, 1, 2$, 
and 4 (the thinnest curve), 
when the multiple SLE in the hydrodynamic limit
starts from the two sources,
$\mu_0=(\delta_1+\delta_{-1})/2$.
As $t \to \infty$, $\partial K_t$ seems to approach
to $\sqrt{t} \partial \cK$.
}
\label{fig:hydro_SLE_double1}
\end{figure}

Figure \ref{fig:hydro_SLE_double1} shows
time dependence of 
the boundary of the SLE hull $\partial K_t$
in the hydrodynamic limit, 
when $a=1$ and 
$\mu_0=(\delta_1+\delta_{-1})/2$.
When $t < t_{\rm c}(1)=1/4$, $\partial K_t$ consists of
two separated curves, which are symmetric
with respect to the imaginary axis. 
These two curves coalesce
at the critical time $t=t_{\rm c}(1)=1/4$,
and then $\partial K_t$ grows as a single curve
in $\H$.

\subsection{$a^2/t$-expansion} \label{sec:expansion}

The numerical plots of $\partial K_t$ for various $t$
in Fig.\ref{fig:hydro_SLE_double1} show that
the double-peak structure in $\partial K_t$ 
disappears gradually as $t \to \infty$ and the
SLE hull $K_t$ seems to approach to a dilatation by factor $\sqrt{t}$ 
of the shape $\cK$ shown in Fig.\ref{fig:hydro_SLE_single1}.
Here we clarify such a long-term asymptotic in $t/a^2 \to \infty$
of $K_t$ starting from the two sources, $\mu_0=(\delta_a+\delta_{-a})/2$,
$a >0$.

We consider the case $t > a^2/4$, in which 
$\supp \mu_t$ is a single interval and is parametrized as
\begin{equation}
\sigma_t(\xi)=\xi a b_+(t/a^2), 
\quad \xi \in I \equiv [-1, 1].
\label{eqn:sigma_2a}
\end{equation}
When we include $a>0$, Eq.(\ref{eqn:ht_eq_double3}) should be read as
\[
\sigma_t(\xi)=h_t(\Gamma_t(\xi))+ \frac{4t h_t(\Gamma_t(\xi))}
{h_t(\Gamma_t(\xi))^2-a^2},
\]
which is rewritten as
\begin{equation}
\frac{1}{t} h_t(\Gamma_t(\xi))^2
-\frac{1}{t} \sigma_t(\xi) h_t(\Gamma_t(\xi))+4
= \left( 1 -\frac{\sigma_t(\xi)}{h_t(\Gamma_t(\xi))} \right)
\frac{a^2}{t}.
\label{eqn:h_s_2a}
\end{equation}
By (\ref{eqn:bpm1}), (\ref{eqn:sigma_2a}) has the following
expansion,
\begin{equation}
\sigma_t(\xi)=\sigma^{(0)} \left\{
1+ \sigma^{(1)} \frac{a^2}{t} + \rO ((a^2/t)^2) \right\}
\label{eqn:sigma_2b}
\end{equation}
with
\begin{equation}
\sigma^{(0)}=\sigma^{(0)}(t, \xi)=4 \xi \sqrt{t},
\quad
\sigma^{(1)}=\frac{1}{8}.
\label{eqn:sigma_2c}
\end{equation}
Then (\ref{eqn:h_s_2a}) implies that $h_t(\Gamma_t(\xi))$ 
can be also expanded with respect to $a^2/t$ as
\begin{equation}
h_t(\Gamma_t(\xi))=h^{(0)} \left\{
1+ h^{(1)} \frac{a^2}{t} + \rO ((a^2/t)^2) \right\}.
\label{eqn:h_exp_1}
\end{equation}
When we put (\ref{eqn:sigma_2b}) and (\ref{eqn:h_exp_1}) into 
(\ref{eqn:h_s_2a}), the 0-th order terms and the first order terms
in the $a^2/t$-expansion give the following equations, respectively,
\begin{align}
& (h^{(0)})^2 - \sigma^{(0)} h^{(0)} + 4t =0,
\label{eqn:exp1}
\\
& 2 (h^{(0)})^2 h^{(1)} - \sigma^{(0)} h^{(0)} ( \sigma^{(1)} + h^{(1)} )
= t \left( 1 - \frac{\sigma^{(0)}}{h^{(0)}} \right).
\label{eqn:exp2}
\end{align} 
Equation (\ref{eqn:exp1}) is the same as (\ref{eqn:eq_single1}), and hence
we have
\[
h^{(0)}=h^{(0)}(t, \xi)=2 \sqrt{t} ( \xi+i \sqrt{1-\xi^2}).
\]
Then (\ref{eqn:exp2}) is solved as
\[
h^{(1)} = h^{(1)}(\xi)=
\frac{(h^{(0)})^2 \sigma^{(0)} \sigma^{(1)} + t (h^{(0)}-\sigma^{(0)})}
{(h^{(0)})^2 (2h^{(0)}-\sigma^{(0)})}
= \frac{3 \xi - i \sqrt{1-\xi^2}}{8(\xi+i \sqrt{1- \xi^2)}}.
\]
If we use the parameterization (\ref{eqn:parameter_varphi1}),
we obtain the following expressions,
\begin{align}
h^{(0)} =& \widetilde{h}^{(0)}(t, \varphi) = 2 i \sqrt{t} e^{-i \varphi},
\nonumber\\
h^{(1)} =& \widetilde{h}^{(1)}(\varphi) = - \frac{i}{8} e^{i \varphi} 
(3 \sin \varphi-i \cos \varphi), \quad
\varphi \in [-\pi/2, \pi/2].
\label{eqn:h0_1}
\end{align}

Now we put (\ref{eqn:h_exp_1}) into
\begin{align*}
\Gamma_t(\xi) =& a V_{t/a^2}(h_t(\Gamma_t(\xi))/a)
\nonumber\\
=& \sqrt{a^2 + (h_t(\Gamma_t(\xi))^2-a^2)
\exp \left( \frac{4 t h_t(\Gamma_t(\xi))^2}{(h_t(\Gamma(\xi))^2-a^2)^2} \right) }.
\end{align*}
We obtain the expansion
\begin{align*}
& \Gamma_t(\xi) = h^{(0)} e^{2t/(h^{(0)})^2}
\nonumber\\
&\quad \times \left[ 1 +
\left\{
\frac{2(h^{(0)})^2 h^{(1)} -t}{2 (h^{(0)})^2}
+ \frac{4t \{t-(h^{(0)})^2 h^{(1)}\}}{(h^{(0)})^4}
+\frac{t}{2(h^{(0)})^2} e^{-4t/(h^{(0)})^2} \right\} \frac{a^2}{t}
+\rO((a^2/t)^2) \right].
\end{align*}
Using (\ref{eqn:h0_1}), we will arrive at the result,
\begin{align*}
\widetilde{\Gamma}_t(\varphi)
\equiv& \Gamma_t( \sin \varphi)
\nonumber\\
=& 2 i \sqrt{t} \exp \left( -i \varphi - \frac{e^{2 i \varphi}}{2} \right)
\left[ 1 + \frac{1}{8} \left\{
1-\exp \left( 2 i \varphi + e^{2 i \varphi} \right) \right\}
\frac{a^2}{t} + \rO((a^2/t)^2) \right],
\nonumber\\
& \hskip 7cm
\varphi \in [-\pi/2, \pi/2]. 
\end{align*}
Therefore, the SLE hull behaves as
\begin{align*}
K_t =& \sqrt{t} \cK (1+\rO(a^2/t) )
\nonumber\\
\equiv& \{ \sqrt{t} z (1+ \rO(a^2/t)) : z \in \cK \}, \quad
\mbox{in $a^2/t \to 0$},
\end{align*}
where $\cK$ is given by (\ref{eqn:SLEhull_universal}).
It implies that for any $a >0$,
\begin{equation}
\lim_{t \to \infty} \frac{K_t}{\sqrt{t}} = \cK
\label{eqn:double_asymptotics}
\end{equation}
for the SLE hull in the hydrodynamic limit
starting from the two sources
$\mu_0=(\delta_a+\delta_{-a})/2$.

In Section \ref{sec:asym} we will discuss that the asymptotic
behavior (\ref{eqn:double_asymptotics}) is universal
for any $\mu_0$ such that
$\supp \mu_0$ on $\R$ is bounded from
either side by constants.

\subsection{Critical curve $\partial K_{t_{\rm c}}$} \label{sec:critical}

\begin{figure}
\begin{center}
\includegraphics[width=1.0\textwidth]{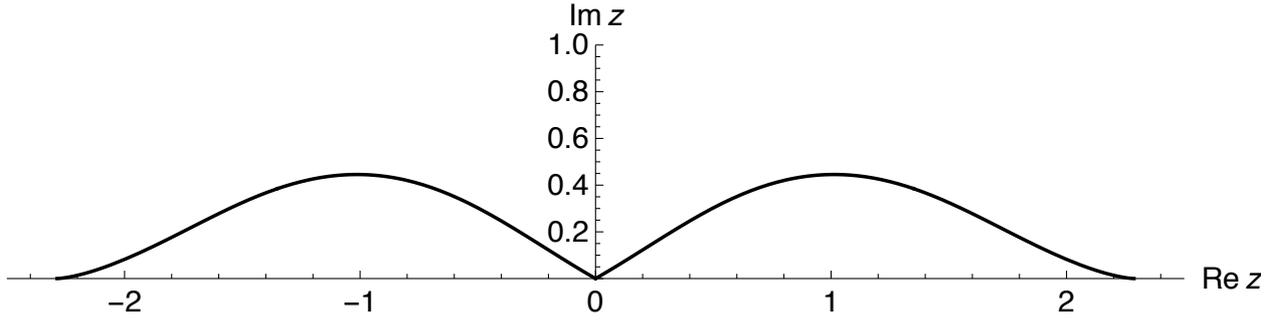}
\end{center}
\caption{
The boundary of the SLE hull in the hydrodynamic limit
at $t=t_{\rm c}=1/4$,
$\partial K_{t_{\rm c}}$, when the system starts
from the two sources, $\mu_0=(\delta_1+\delta_{-1})/2$.
The curve is symmetric with respect to the imaginary axis.
The three osculation points on $\R$ are located at
0 and $\pm x_{\rm c}=\pm \sqrt{1+2e^{3/4}}=\pm 2.287 \cdots$.
At the edges $x=\pm x_{\rm c}$, the curve $\partial K_{t_{\rm c}}$ shows
the power law $\sim |x \mp x_{\rm c}|^{3/2}$.
In the vicinity of the origin, the curve
is $V$-shaped with an inner angle $2 \pi/3$.
}
\label{fig:critical_shape}
\end{figure}

Here we assume $a=1$ for simplicity of expressions.
At the critical time $t_{\rm c}=t_{\rm c}(1)=1/4$,
(\ref{eqn:bpm1b}) gives
\[
b_-(t_{\rm c})=0, \quad
b_+(t_{\rm c})=\frac{3 \sqrt{3}}{2}.
\]
Let
\[
\sigma_{t_{\rm c}}(\xi)= \xi b_+(t_{\rm c})
=\frac{3 \sqrt{3}}{2} \xi, \quad \xi \in [-1, 1],
\]
which parameterizes the whole interval of 
$\supp \mu_{t_{\rm c}}=[-3 \sqrt{3}/2, 3 \sqrt{3}/2]$.
The equations to determine the boundary of the
SLE hull 
\[
\partial K_{t_{\rm c}}=
\{ \Gamma_{t_{\rm c}}(\xi)=V_{t_{\rm c}}(v+i w) :
v=v(\xi) \in \R, w =w(\xi) \in \R, 
\xi \in [-1, 1] \}
\]
are given by
\begin{align}
& 8 v^3-12 \sqrt{3} \xi v^2 + \frac{27}{2} \xi^2 v
-\frac{3 \sqrt{3}}{2} \xi =0,
\nonumber\\
& w^2= 3 v(v-\sqrt{3} \xi),
\label{eqn:vw_c1}
\end{align}
where
\begin{equation}
V_{t_{\rm c}}(z)=V_{1/4}(z)
=\sqrt{1+(z^2-1) e^{z^2/(z^2-1)^2} }.
\label{eqn:Vc1}
\end{equation}
Figure \ref{fig:critical_shape} shows the
critical curve $\partial K_{t_{\rm c}}$,
which is symmetric with respect to the imaginary axis.

If we set $\xi=\pm 1$ and $w=0$,
the equations (\ref{eqn:vw_c1}) become
\begin{align*}
& (v \mp \sqrt{3}) \left( 8 v^2 \mp 4\sqrt{3} v+ \frac{3}{2} \right) =0,
\nonumber\\
& 3 v (v \mp \sqrt{3} ) =0.
\end{align*}
Then we find the solution 
\[
h_{t_{\rm c}}(\Gamma_{t_{\rm c}}(\pm 1))
= \pm \sqrt{3} \equiv \pm v_{\rm c}. 
\]
On the other hand, if we set $\xi=0$ and $w=0$,
the equations (\ref{eqn:vw_c1}) give
\[
h_{t_{\rm c}}(\Gamma_{t_{\rm c}}(0))=0.
\]
They determine the three real values in 
the curve $\partial K_{t_{\rm c}} \in \overline{\H}$, 
\begin{align*}
\Gamma_{t_{\rm c}}(\pm 1)
=& \pm V_{t_{\rm c}}(\pm v_{\rm c})
= \pm \sqrt{1+2 e^{3/4}}
\nonumber\\
=& \pm 2.287 \cdots \equiv x_{\rm c},
\nonumber\\
\Gamma_{t_{\rm c}}(0)
=& V_{t_{\rm c}}(0)=0,
\end{align*}
which are the three osculation points
of $\partial K_{t_{\rm c}}$ on $\R$;
$\partial K_{t_{\rm c}} \cap \R
=\{-x_{\rm c}, 0, x_{\rm c}\}$.

If we put $z=v+\delta$ in (\ref{eqn:Vc1}) with
$v \in \R, \delta \in \C$,
$v \not=0, |\delta| \ll 1$,
we have the following expansion,
\begin{align*}
V_{t_{\rm c}}(v+\delta)
=& \sqrt{1+(v^2-1) e^{v^2/(v^2-1)^2} }
\nonumber\\
&
\times \left[
1+ \frac{e^{v^2/(v^2-1)^2}}{1+(v^2-1) e^{v^2/(v^2-1)^2}}
\frac{v^3(v^2-3)}{(v^2-1)^2} \delta
\right.
\nonumber\\
& \quad
+ \frac{e^{v^2/(v^2-1)^2}}{1+(v^2-1) e^{v^2/(v^2-1)^2}}
\frac{v^2(v^8-6v^6+18v^4-14v^2+9)}{2(v^2-1)^5} \delta^2
\nonumber\\
& \quad \left.
- \frac{e^{2 v^2/(v^2-1)^2}}{ \{1+(v^2-1) e^{v^2/(v^2-1)^2} \}^2}
\frac{v^6(v^2-3)^2}{2(v^2-1)^4} \delta^2 +
\rO(\delta^3) \right].
\end{align*}
In the vicinity of the right edge $x=x_{\rm c}$ of $\partial K_{t_{\rm c}}$,
the above gives
\begin{equation}
V_{t_{\rm c}}(v_{\rm c}+\delta)
=x_{\rm c} +\frac{9 e^{3/4}}{4 x_{\rm c}} \delta^2 + \rO(\delta^3).
\label{eqn:Vc3}
\end{equation}
If we set $\xi=1-\varepsilon$,
$v=v_{\rm c}+c \varepsilon +\rO(\varepsilon^2)$
in (\ref{eqn:vw_c1}) assuming $0 < \varepsilon \ll 1$,
the coefficient $c$ is determined as
$c=-7 \sqrt{3}/9$ and we obtain
$w=\pm \sqrt{2} \varepsilon^{1/2}+\rO(\varepsilon^{3/2})$.
We put
\[
\delta=-\frac{7 \sqrt{3}}{9} \varepsilon - \sqrt{2} i \varepsilon^{1/2}
+\rO(\varepsilon^{3/2})
\]
in (\ref{eqn:Vc3}). Then we have
\[
V_{t_{\rm c}}(v_{\rm c}+\delta)
=x_{\rm c} 
-\frac{9 e^{3/4}}{2 x_{\rm c}} \varepsilon 
+ \frac{7 \sqrt{6} e^{3/4}}{2 x_{\rm c}} i \varepsilon^{3/2}
+\rO(\varepsilon^2).
\]
The above result implies that 
in the vicinity of the right and left edges $\pm x_{\rm c}$, 
the boundary of the SLE hull,
$\{z=x+iy : z \in \partial K_{t_{\rm c}} \}$, behaves as
\[
y \simeq \begin{cases}
\displaystyle{ \frac{14 \sqrt{6}}{27} \sqrt{\frac{x_{\rm c}}{x_{\rm c}^2-1}}  
(x_{\rm c} - x)^{3/2} },
\quad & \mbox{if $x \leq x_{\rm c}$},
\\
\displaystyle{ \frac{14 \sqrt{6}}{27} \sqrt{\frac{x_{\rm c}}{x_{\rm c}^2-1}} 
(x+x_{\rm c})^{3/2} },
\quad & \mbox{if $x \geq - x_{\rm c}$}.
\end{cases}
\]
The singularities of the curve $\partial K_{t_{\rm c}}$ 
at the edges $x=\pm x_{\rm c}$ are governed by
the power law with exponent $3/2$,
which is common to the single-source solution 
as shown by (\ref{eqn:edges_single}). 

Next we consider the vicinity of the origin.
Equation (\ref{eqn:Vc1}) gives
\begin{equation}
V_{t_{\rm c}}(\delta)=
\sqrt{ - \frac{3}{2} \delta^4 + \rO(\delta^6) },
\quad |\delta| \ll 1.
\label{eqn:Vc5}
\end{equation}
If we assume $0 < \xi \ll 1$, $|v| \ll 1$ and $|w| \ll 1$,
then the function (\ref{eqn:vw_c1}) behaves as
\[ 
v = v(\xi) \simeq \frac{\sqrt{3}}{2^{4/3}} \xi^{1/3}, \quad
w = w(\xi) \simeq \sqrt{3} v(\xi).
\]
Then if we put
\begin{align*}
\delta =& \delta(\xi)  = v(\xi) + iw(\xi)
\nonumber\\
\simeq& \frac{\sqrt{3}}{2^{4/3}} (1+ i \sqrt{3}) \xi^{1/3}
=\frac{\sqrt{3}}{2^{1/3}} e^{i \pi/3} \xi^{1/3}
\end{align*}
in (\ref{eqn:Vc5}), we obtain the estimate
\[
\Gamma_{t_{\rm c}}(\xi)
=V_{t_{\rm c}}(\delta(\xi))
\simeq \sqrt{ \frac{3^3}{2^{7/3}} e^{i \pi /3} \xi^{4/3} }
=\frac{3^{3/2}}{2^{7/6}} e^{i \pi/6} \xi^{2/3},
\quad 0 < \xi \ll 1.
\]
This result implies that
in the vicinity of the origin,
the boundary of the SLE hull,
$\{z=x+iy : z \in \partial K_{t_{\rm c}}\}$, behaves as
\[
y \simeq \pm \frac{x}{\sqrt{3}}.
\]
That is, $\partial K_{\rm t_{\rm c}}$ behaves linearly
in the vicinity of the origin, and
the right and left asymptotic lines make angles 
with respect to the positive real axis 
given by $\pi/6$ and $\pi-\pi/6$, respectively.

Note that after the critical time, $t > t_{\rm c}$,
the singularity at the origin disappears 
in the curve $\partial K_t$, since 
for $t > t_{\rm c}$ there is no gap in
$\supp \mu_t$ as given by (\ref{eqn:supp_d1}) 
and the map between $\xi \in \supp \mu_t$
and $\Gamma_t(\xi) \in \partial K_t$ is analytic
as shown by (\ref{eqn:ht_eq_double3}).

\SSC
{Long-Term Asymptotics} \label{sec:asym}
\subsection{Wigner's semicircle law as a long-term asymptotics} \label{sec:Wigner}

For a function of $t$ and $z$, $f(t,z)$, here we use the
following abbreviations for partial differentials,
\[
\dot{f} =\frac{\partial f}{\partial t}, \quad
f'= \frac{\partial f}{\partial z}.
\]
Consider the complex Burgers equation (\ref{eqn:Mt2}) 
for $M_t(z)$, $t \geq 0, z \in \H$,
which is now written as
\begin{equation}
\dot{M}_t(z)=-2M_t(z) M_t'(z).
\label{eqn:dM2}
\end{equation}
Let $c>0$ and define
\begin{equation}
M_t(z,c) \equiv c M_{c^2 t}(cz).
\label{eqn:Mtc}
\end{equation}
We see that
\[
\frac{\partial M_t(z,c)}{\partial t}
= c^3 \dot{M}_{c^2t}(cz),
\quad
\frac{\partial M_t(z, c)}{\partial z}
= c^2 M_{c^2t}'(cz).
\]
If we set $t \to c^2t, z \to cz$ in (\ref{eqn:dM2}), we have the equation
\[
\dot{M}_{c^2t}(cz)=-2 M_{c^2t}(cz) M_{c^2t}'(cz).
\]
Multiply the both sides by $c^3$. Then we obtain the equation
\[
\frac{\partial M_t(z,c)}{\partial t}
=-2 M_t(z,c) \frac{\partial M_t(z,c)}{\partial z}.
\]
Hence, 
if $M_t(z)$ solves the complex Burgers equation (\ref{eqn:dM2}), then
(\ref{eqn:Mtc}) with any $c>0$ also solves it.

We find that 
\begin{align*}
M_0(z,c) &= c M_0(cz)
\nonumber\\
&= c \int_{\R} \frac{2 \mu_0(du)}{cz-u}
=\int_{\R} \frac{ 2 \mu_0(du)}{z-u/c},
\end{align*}
and hence, if the support of $\mu_0$ is bounded from either
side by constants, 
\begin{equation}
\supp \mu_0 \subset [-L, L] \quad \mbox{for some $L>0$}, 
\label{eqn:bounded_support}
\end{equation}
then
\[
\lim_{c \to \infty} M_0(z,c)
= \frac{2}{z} \int_{\R} \mu_0(du) = \frac{2}{z}.
\]
Therefore, for any fixed $T < \infty$, 
if (\ref{eqn:bounded_support}) is satisfied,
\begin{equation}
\lim_{c \to \infty} M_T(z,c)
= \lim_{c \to \infty} c M_{c^2 T}(cz)
=\frac{4}{z+\sqrt{z^2-16 T}}
\label{eqn:MT1}
\end{equation}
is concluded by the result (\ref{eqn:Mt_single}).

Let ${\bf 1}(\omega)$ be an indicator function;
${\bf 1}(\omega)=1$ if the condition $\omega$ is satisfied,
and ${\bf 1}(\omega)=0$ otherwise.
We can prove the following 
for the hydrodynamic limit of the Dyson model.

\begin{prop}
\label{thm:Mt_asym}
For any initial distribution
$\mu_0$ satisfying (\ref{eqn:bounded_support}), 
\begin{align}
\label{eqn:Mt_asym1}
& \lim_{t \to \infty} \sqrt{t} M_t(\sqrt{t} z)
=\frac{4}{z+\sqrt{z^2-16}},
\\
\label{eqn:rhot_asym1}
& \lim_{t \to \infty} \sqrt{t} \mu_t(\sqrt{t} du)
= {\bf 1}(|u| \leq 4)
\frac{1}{8 \pi} \sqrt{16-u^2} du.
\end{align}
\end{prop}
\noindent{\it Proof.} \,
In (\ref{eqn:MT1}), if we set
$c=\sqrt{t/T}$,
we have
\[
\lim_{t \to \infty} \sqrt{ \frac{t}{T} }
M_t \left( \sqrt{\frac{t}{T}}  z \right)
=\frac{4}{z+\sqrt{z^2-16T}}.
\]
Since $T$ is an arbitrary positive number,
we can replace $z/\sqrt{T}$ by $z$ and obtain (\ref{eqn:Mt_asym1}). 
On the other hand, as given by (\ref{eqn:Wigner}), 
for $\mu_0=\delta_0$,
\begin{align*}
M_t(z) &= \frac{4}{z+\sqrt{z^2-16t}}
\nonumber\\
&=\int_{\R} \frac{2}{z-u} {\bf 1}(|u| \leq 4 t)
\frac{1}{8 \pi t} \sqrt{16t-u^2} du.
\end{align*}
If we set $t=1$, the above implies that  
\[
\mbox{(RHS) of Eq.} (\ref{eqn:Mt_asym1}) = 
\int_{\R} du \, \frac{2}{z-u}
{\bf 1}(|u| \leq 4) \frac{1}{8 \pi} \sqrt{16-u^2}.
\]
By the definition of Stieltjes transformation (\ref{eqn:Mt1}), 
we see that
\[
\mbox{(LHS) of Eq.} (\ref{eqn:Mt_asym1})
= \lim_{t \to \infty} \sqrt{t}
\int_{\R} \frac{2 \mu_t(du)}{\sqrt{t} z-u}
= \lim_{t \to \infty}
\int_{\R} du \,
\frac{2 \mu_t(du)}{z-u/\sqrt{t}}.
\]
If we change the integral variable as 
$u \to w \equiv u/\sqrt{t}$, the above is written as
\[
\lim_{t \to \infty} \int_{\R} 
\frac{2 \sqrt{t} \mu_t(\sqrt{t} dw)}{z-w}.
\]
Then (\ref{eqn:rhot_asym1}) is concluded. 
The proof is hence completed. 
\qed

\subsection{Long-term asymptotics of hydrodynamic limit of 
the multiple SLE} \label{sec:long_t_SLE}

The statement (\ref{eqn:Mt_asym1}) in
Proposition \ref{thm:Mt_asym}
means the asymptotics
\begin{equation}
M_t(z) \simeq \frac{1}{\sqrt{t}}
\frac{4}{z/\sqrt{t}+\sqrt{(z/\sqrt{t})^2-16}}
=\frac{4}{z+\sqrt{z^2-16t}},
\quad \mbox{in $t \to \infty$}.
\label{eqn:Mt_asym3}
\end{equation}
This property was stated as Remark 2.13 in \cite{dMHS17}.
Here we claim that, by the consideration given in 
Section \ref{sec:single_source_SLE},
(\ref{eqn:Mt_asym3}) implies that
\[
g_t(z) \simeq 2 i \sqrt{t} \left\{
\frac{1}{\sqrt{W_0(-4t/z^2)}}-\sqrt{W_0(-4t/z^2)} \right\}, 
\quad \mbox{in $t \to \infty$},
\]
and hence 
\[
K_t \simeq \sqrt{t} \cK,
\quad \mbox{in $t \to \infty$}, 
\]
where $\cK$ is given by (\ref{eqn:SLEhull_universal}). 
Therefore, the following theorem will be established.
\begin{thm}
\label{thm:SLE_asym}
For any $\mu_0$ satisfying (\ref{eqn:bounded_support}),
the hydrodynamic limit of 
the multiple SLE shows the following
long-term asymptotics,
\[
\lim_{t \to \infty} \frac{1}{\sqrt{t}} g_t(\sqrt{t} z)
= 2i \left\{
\frac{1}{\sqrt{W_0(-4/z^2)}}-\sqrt{W_0(-4/z^2)}
\right\},
\quad z \in \overline{\H} \setminus \cK, 
\]
where $\cK$ is given by (\ref{eqn:SLEhull_universal}). 
\end{thm}

\SSC
{Concluding Remarks} \label{sec:remarks}

For $N \in \N$, the $N$-tuple SLE is a coupled system of 
(a modification of) the Dyson model $\V^N(t)$
driven by a set of independent Brownian motions
$\B(t)=(B_1(t), \dots, B_N(t))$ on $\R$
and the differential equation of conformal map $g^N_t$ onto $\H$
driven by $\V^N(t)$ \cite{BBK05,KL07,dMS16,dMHS17}.
The probability law of solution $g^N_t$ 
governs the statistical ensemble of $N$ random-slits in $\H$.
Recently del Monaco and Schlei{\ss}inger \cite{dMS16} 
discussed the system in the limit $N \to \infty$.
In the case of simultaneous growth of slits in $\H$
without any intersection, they showed that
the limit is a deterministic system consisting of
a $(1+1)$-dimensional partial differential equation
called the complex Burgers equation (\ref{eqn:Mt2})
for $M_t(z), t \geq 0, z \in \C \setminus \R$,
and the ordinary equation for $g_t$ (\ref{eqn:dg})  
driven by $M_t(\cdot)$.
Since the complex Burgers equation has been studied
in discussing the hydrodynamic limit of the Dyson model
\cite{Cha92,RS93,AGZ10,BN10,FG15,AK16}, here 
we have regarded the limit system as the
hydrodynamic limit of the multiple SLE.

The Dyson model is a dynamical extension of the
eigenvalue statistics studied in random matrix
theory \cite{Dys62,Meh04,For10,AGZ10,Kat15_Springer}.
The most fundamental probability-law
for eigenvalues of random matrices is
Wigner's semicircle law,
since it can be considered as 
the {\it law of large numbers} 
for the eigenvalue statistics of random matrices.
The universality of Wigner's semicircle law
has been extensively studied.
See \cite{ESY09a,ESY09b,TV10,TV11} and 
references therein.
It is also well-known that
if we consider the complex Burgers equation
under the initial condition $\mu_0=\delta_0$,
it has a unique solution, which we denote as $M_t^{\rm Wigner}(z)$ here,
since it is identified with the
Stieltjes transformation of the 
measure following the time-dependent version
of Wigner's semicircle law
(see, for instance, Exercise 4.3.18 in \cite{AGZ10}, and \cite{AK16}). 

Following \cite{dMS16}, we gave the conformal map
$g_t(z)$ driven by $M_t^{\rm Wigner}(\cdot)$
using the Lambert function $W_0$ in Proposition \ref{thm:SLE_gt_single}.
We would like to emphasize the importance
of its hull $K_t$ given by Proposition \ref{thm:SLEhull_single},
since it describes the time evolution of
the hydrodynamic limit of an infinite number of
slits growing in $\H$.
As demonstrated in Sections \ref{sec:expansion} and \ref{sec:asym},
we expect that the hull $\cK=K_1$
given by (\ref{eqn:SLEhull_universal}) and Fig.\ref{fig:hydro_SLE_single1}
will provide a {\it universal shape} describing the long-term behavior of systems.

One of the advantage of SLE \cite{Sch00,LSW04,Law05}
is that it enables us to clarify the probability laws of
{\it random fractal-geometry}, {\it e.g.}, the SLE slits
and its hull, by analyzing the stochastic differential equation
for a conformal map.
As Wigner's semicircle distribution plays a role 
of the law of large numbers in random matrix theory,
the shape $\cK$ will be considered to 
represent the law of large numbers 
for the {\it infinite system of interacting 
random curves} in $\H$ generated
by the multiple SLE.

As well as the random matrix theory provides 
a universal viewpoint for statistical and stochastic systems,
it has the great variety of ensembles depending on
symmetry and geometrical restrictions.
In suitable setting, we can also 
observe crossover phenomena \cite{KT04}
and phase transitions with critical phenomena 
\cite{BN08,BN10,FMS11,SMCF13,GMN16}.
It will be an interesting future problem
to clarify how to lift these rich structures
in the random matrix theory up to the level of
multiple SLE.

\vskip 0.5cm
\noindent{\bf Acknowledgements} \quad
This paper is based on the manuscript prepared for
the presentation given in 
`Tokyo-Seoul Conference in Mathematics
-- Probability Theory --', 
which was held in the University of Tokyo,
December 8 and 9, 2017.
The present authors thank the organizers, 
Shigeki Aida, Nam-Gyu Kang, JongHae Keum, and Toshitake Kohno, 
very much for giving them such an opportunity to give a talk.
One of the present authors (I.H.) is supported by
the Grant-in-Aid for Young Scientists (B) (No.17K14205)
of Japan Society for
the Promotion of Science.
Another author (M.K.) is supported by
the Grant-in-Aid for Scientific Research (C) (No.26400405),
(B) (No.26287019), and
(S) (No.16H06338) of Japan Society for
the Promotion of Science.


\end{document}